\documentclass[a4paper]{article}

\usepackage[english]{babel}
\usepackage[utf8x]{inputenc}
\usepackage[T1]{fontenc}

\usepackage[a4paper,top=3cm,bottom=2cm,left=3cm,right=3cm,marginparwidth=1.75cm]{geometry}

\usepackage{amsmath}
\usepackage{graphicx}
\usepackage[colorinlistoftodos]{todonotes}
\usepackage[colorlinks=true, allcolors=blue]{hyperref}

\title{Optical probe pressure effects on cutaneous blood flow}
\author{I. A. Mizeva\\
Institute of Continuous Media Mechanics of the Ural Branch, RAS,  Perm, Russia \\
Elena V. Potapova, Viktor V. Dremin, Evgeny A. Zherebtsov, \\Mikhail A. Mezentsev, Valerii V. Shupletsov, Andrey V. Dunaev\\
Orel State University named after I.S. Turgenev, Orel, Russia
}

\begin{document}
\maketitle

\begin{abstract}
The variation of blood flow characteristics caused by the probe pressure during noninvasive studies is of particular interest within the context of fundamental and applied research. It has been shown previously that the weak local pressure induces vasodilation, whereas the increased pressure is able to stop the blood flow in the compressed area, as well as to significantly change optical signals.

The blood flow oscillations measured by laser Doppler flowmetry (LDF) characterize the functional state of the microvascular system and can be used for noninvasive diagnostics of its abnormality. This study was intended to identify the patterns of the relationship between the oscillating components of  blood flow registered by the LDF method under different levels of pressure applied to an optical fiber probe.

For this purpose we have developed an original optical probe capable of regulating the applied pressure. The developed protocol included six sequential records of the blood perfusion at pressure within the 0 to 200 mmHg  range with unloading at the last stage.

Using wavelet analyses, we traced the variation of energy of oscillations for these records in five frequency bands associated with different vascular tone regulation mechanisms. Six young volunteers of the same age (three males and three females) were included in this preliminary study and the protocol was repeated five times in each volunteer. Totally 30 LDF records were available for the analyses. As expected, the LDF signal increases at weak pressure (30 mmHg) and decreases at increased pressure. The statistically stable amplification of endothelial associated blood flow oscillations under the 90 mmHg pressure allowed us to put forward a hypothesis that the endothelial activity increases. The possible causes of this phenomenon are discussed.
\end{abstract}

\section{Introduction}

The local tissue perfusion is generally regulated by microvascular tone,  which is under control of a number of physiological mechanisms. Together they lead to a sequence of vasoconstrictions and vasodilation which form fluctuating regime of vasomotion. Physiologically these mechanisms are intended to ensure metabolic requirements of the tissue and interrelated with its metabolic state \cite{Jacob2016}.

External factors such as temperature, local pressure of the probe on the skin, etc. influence the blood flow characteristics. Moreover such factors are used in provocative tests to evaluate the functional state of microvascular regulatory mechanisms. The tests allow one to define adaptation reserve of the microvascular network to external stress and are utilized to study microcirculation abnormalities which accompany a variety of pathologies.

From this point of view the effects of local skin pressure 
caused by external pressure are of particular interest due the fact that the local tissue ischaemia, modeled by such an impact, appears to be one of the main determinants of skin lesions in certain clinical situations (pressure sores \cite{Agrawal2012} and diabetic foot ulcers \cite{Mills}). 
In previous studies on this topic, most spectroscopic measurements have been conducted by means of a topical optical fiber probe that transmits and collects optical spectra from the skin. Probe pressure influences the microcirculation parameters which can be measured by optical methods: diffuse reflectance spectroscopy \cite{Lim2011, Popov2017}, fluorescence spectroscopy \cite{Lim2011,Zherebtsov2018, Meglinski2002}  and Laser Doppler flowmetry (LDF) \cite{Zherebtsov2017, Sacks1988}. 

LDF \cite{Stern1975} together with wavelet analyses \cite{Stefanovska1999} is a commonly used technique to study the regulatory mechanisms of cutaneous microcirculation. LDF is based on the extraction information about the parameters of optical Doppler shift of scattered and reflected laser radiation in the living tissue containing moving red blood cells. In a first approximation, the LDF signal is proportional to the number of scattering particles in the volume of the tissue multiplied by their velocity. The method has found many applications in research and clinical practice including the study of perfusion response to the pressure induced vasodilation (PIV) \cite{ABRAHAM2001122}. 

Specifically this parameter has been demonstrated to be sensitive to endothelial dysfunction in diabetes \cite{Sigaudo-Roussel1564} \cite{Schubert89}. Using similar approach, it was demonstrated \cite{Fuyuan2017} that the test based on moderate skin compression can be utilized to study microvascular reactivity and to determine conditions with a risk of diabetic foot ulcer development and progression.

Apart from mean perfusion, skin blood flow oscillations associated with different types of the vascular tone regulation \cite{Mizeva2018} can be measured by the LDF. 

Information about variations in parameters of the oscillations during local skin compression is of particular importance for  fundamental and applied research as it has potential to underlie new diagnostic protocols. However, at present just few researchers address this problem.  For example, the variation of skin blood flow oscillations caused by local pressure allowed Y.K. Jan \cite{Jan2012, JAN2008} to study the influence of factors that decrease ischemia in weight-bearing soft tissues, which prevents pressure ulcers. An LDF time-frequency analysis  provides additional information to ensure better understanding of the PIV phenomenon \cite{Humeau2004}.

The aim of this work is to identify patterns of the relationship between the oscillating components of  blood flow registered by the LDF method under different levels of pressure applied to the skin.

We put particular emphasis on studying the inter-subject and inter-group variability of LDF spectral characteristics and variations in oscillating components caused by  local skin compression. Our ultimate goal is to reveal the distinguishing features of the  obtained patterns, which can be used in clinical studies.

\section{Materials and Methods}
\label{sec:MM}

\subsection{Protocol design}
 
\begin{figure}
\begin{center}
\begin{tabular}{c}
\includegraphics[height=7.5cm]{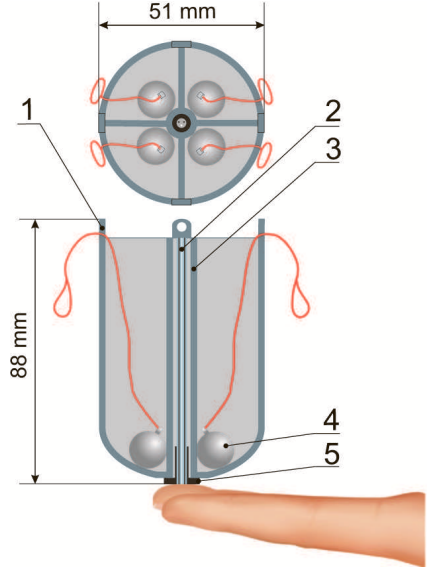}
\end{tabular}
\end{center}
\caption { Scheme of the experimental setup. 1 - pressure distribution tool, 2 - LDF probe, 3 - parallel hole for the LDF probe, 4 - leads, 5 - tip of the LDF probe.}
\label{fig:setup}
\end{figure}

The measurements were carried out by the laser Doppler flowmeter LAKK-02 (SPE LAZMA Ltd., Russia) with laser probing radiation of 1064 nm. The calibrated pressure was applied to skin by use of an in-house designed tool (1 in the Fig.\ref{fig:setup}), \cite{Zherebtsov2017,Zherebtsov2018} which allowed one to distribute the weight of placed leads (4) over the LDF probe (2) installed into the parallel hole (3). During the experiment the probe tip (5) was the only area of contact with skin. The contact area was of 70 mm$^{2}$.

During the measurement procedure, the subject was sitting in a relaxed condition. His right hand was placed on the table at heart level. An LDF probe was located on the dorsal surface of the distal phalanx of the third finger. In our investigation we used a stepwise loading protocol, and the duration of each LDF record after every change of pressure was 10 minutes. First, we collected LDF records in  basal conditions and then increased the pressure applied to the probe in stepwise fashion. The pressure values were as follows: 30 (stage 2), 90  (stage 3),  140 (stage 4), 200 (stage 5) and 30 mmHg (stage 6). It should be noted that at the last stage the leads were removed and the skin compression was the same as at  stage 2 (due to the proper weight of the tool and the optical probe). In comparison to normal blood pressure values, the chosen pressures correspond to a diastolic blood pressure (DBP) that is lower than normal DBP, slightly higher than normal DBP, slightly higher than normal systolic blood pressure (SBP) and much higher than SBP, respectively.

\subsection{Subjects}
   
Six volunteers (three males and three females) were involved in the study. The average age was $20\pm2$ years in both groups. The subjects were caffeine and medications free. To avoid movement artifacts, the volunteers were asked to minimize speaking or moving during the measurement session.  All of the subjects gave the signed and informed consent before the experiment. In this study we collected 5 records from every subject in the group. Accordingly 30 LDF samples were available for the analyses. The study was approved by the Ethics Committee at the Orel State University named after I.S. Turgenev. 

\subsection{Data analysis}

The LDF samples ($f(t)$) are post-processed by means of the original wavelet decomposition software \cite{Mizeva2016, Mizeva2017JBO}. Thus we have 

\begin{equation}
W(\nu,\tau)=\nu\int\limits_{-\infty}^{\infty}f(t)\psi^{*}(\nu(t-\tau))dt,
\label{CWT}
\end{equation}
where the symbol $^*$ denotes complex conjugation, $\nu$ is the frequency, $f(t)$ is the sample under consideration, $t$ is the time, and $\tau$ is the time shift. The Morlet wavelet written in the form
\begin{equation}
\psi(t)=e^{2\pi i t}e^{-t^2/\sigma} \label{morlet}
\end{equation}
is used in a series expansion with the decay parameter $\sigma=1$.  Such analyzing function allows one to ensure sufficient time-frequency resolution and, being localized in the time-space, it provides one with better statistics for the slowest oscillations. In the first step, we calculate the wavelet coefficients for the whole LDF sample, then separately integrate the power at each step ($i$) taking into account boundary effect by step away $\Delta$ from every step interval boundaries $T_i$ : 
\begin{equation}
M(\nu)=\frac{1}{T}\int\limits_{T_i+\Delta}^{T_{i+1}-\Delta} |W(\nu,t)|^2 dt.  \label{eq:spec}
\end{equation}

Also, we consider the spectral properties in five frequency bands associated with different mechanisms of vascular tone regulation. The cardiac bands "C" (0.45-1.6 Hz) and the respiratory (0.2-0.45 Hz) bands "R" carry information about the influence of heart rate and movement of the thorax on the peripheral blood flow. The myogenic ("M") mechanism of vascular tone regulation reflects the response of vascular smooth muscle cells to the transmural pressure. Blood flow oscillations at frequencies (0.05-0.15 Hz) characterize its activity. The neurogenic sympathetic vasomotor activity ("N") causes the vessel walls to move with frequency 0.02-0.05 Hz. Slow blood flow waves (0.005-0.0095 Hz and 0.0095-0.02 Hz) reflect the vascular tone regulation due to the endothelium activity ("E"), both NO-dependent and independent. These mechanisms were reviewed in detail \cite{Kvernmo1999, Lancaster2015}.

The statistical analysis was performed with Mathematica 8.0. 

\section{Results}
\label{sec:results}

\begin{figure}
\begin{center}
\begin{tabular}{c}
\includegraphics[height=2.8cm]{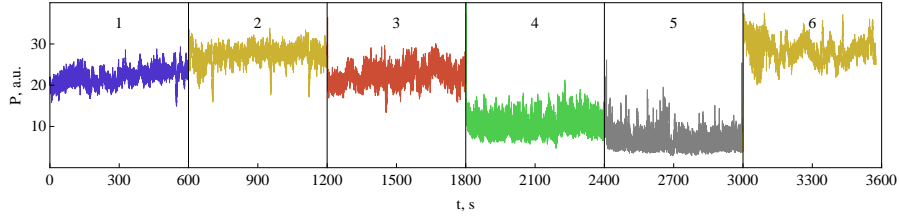}
\end{tabular}
\end{center}
\caption
{ \label{fig:LDFExample}
Typical example of the LDF sample in accordance with the protocol used in this study; different colors and numbers 1..6 indicate different stages of the experiment, which are separated by vertical lines}
\end{figure}

A typical example of the LDF sample is shown in Fig.~\ref{fig:LDFExample}. The local pressure induces redistribution of perfusion. In the second stage of the experiment (30 mmHg), perfusion first increases and then, under  subsequent loading, gradually decreases until the moment of unloading, when a splash corresponding to reactive hyperemia is recorded (as shown on the graph). Further we quantify  a variation in the mean value of perfusion caused by different applied pressures as $\delta P_i=P_i-P_1$, where the subscript $i$ indicates the test number ($i=2..6$), and $P_1$ is the mean basal perfusion for the experiment. 

\begin{figure}
\begin{center}
\begin{tabular}{c}
\includegraphics[height=4.5cm]{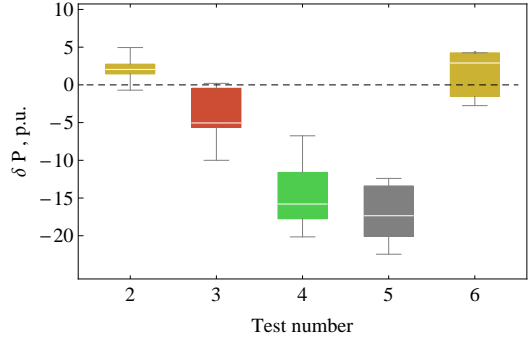}
\end{tabular}
\end{center}
\caption
{ \label{fig:MeanSt}
Group-averaged variation in the mean values of perfusion caused by local loading. }
\end{figure}

Intra- and intersubject reproducibility of the mean perfusion variation $\delta P_i$ (Fig.~\ref{fig:MeanSt}) is assessed. The low pressure ($i=2$, 30 mmHg) induces vasodilation ($\delta P > 0$), and further loading induces vasoconstriction ($\delta P < 0$). Note that a decrease in perfusion is attributed to the external impact but not to the internal physiological mechanisms of vasoconstriction.
The increase in the skin blood flow observed in the second load period is, probably, associated with PIV and may also be caused by tissue deformation and redistribution of blood with a change in the geometry of the vessels in the diagnostic volume. 

\begin{figure}
\begin{center}
\begin{tabular}{c}
\includegraphics[height=3.7 cm]{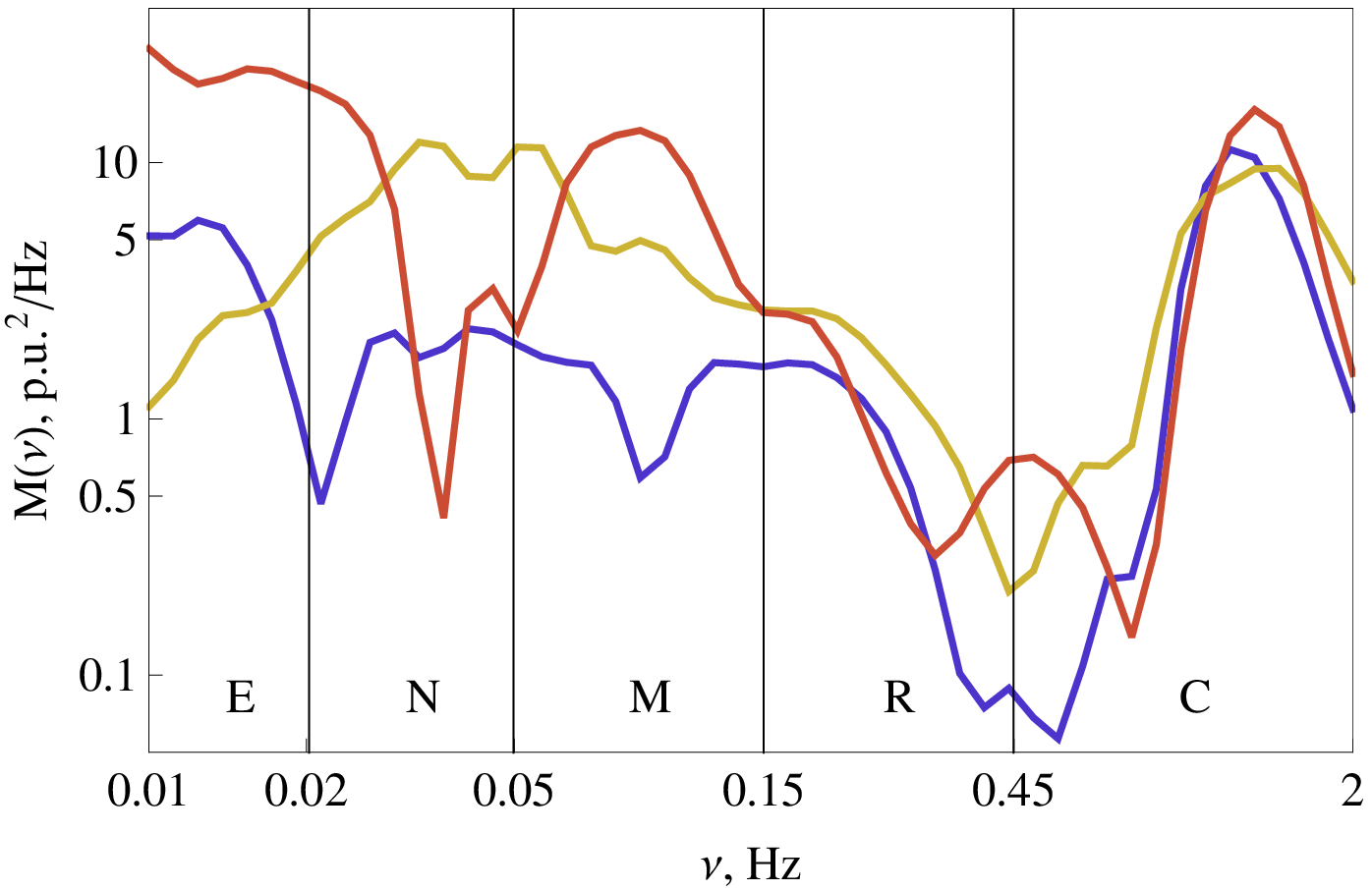}
\includegraphics[height=3.7 cm]{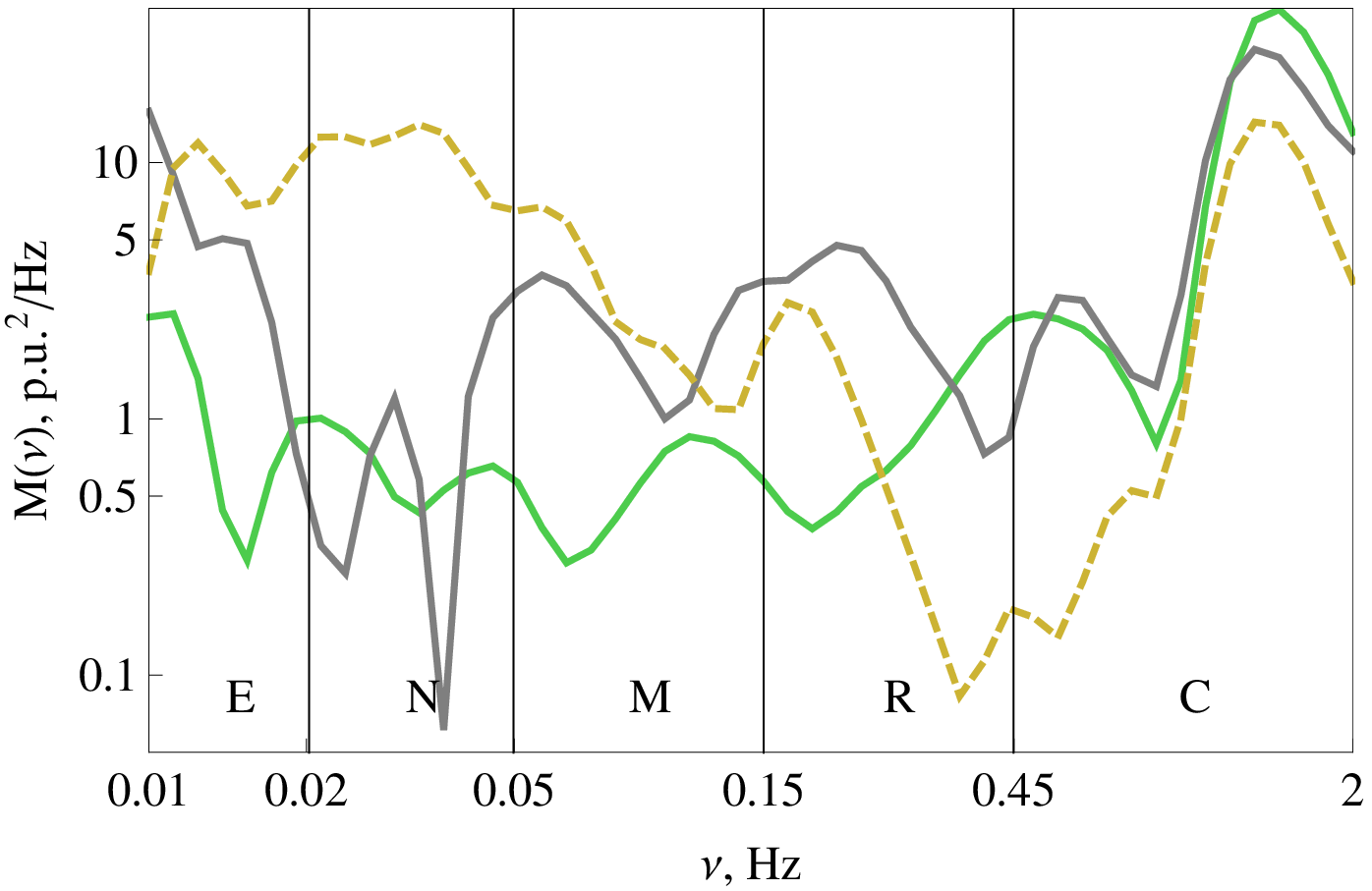}
\end{tabular}
\end{center}
\caption
{ \label{fig:SpExamp}
Spectra of the LDF sample are presented in  Fig.~\ref{fig:LDFExample}. Colors are the same as in the  previous figures. Left panel - stages 1-3 of the experiment (blue -- basal, yellow -- 30 mmHg, and red -- 90 mmHg loading), right panel -- further stages of the experiment (green -- 140 mmHg, gray -- 200 mmHg, and dashed yellow -- 30 mmHg --  final stage (6) after taking the load off.  } 
\end{figure}

\begin{figure}
\begin{center}
\begin{tabular}{c}
\includegraphics[height=3.7 cm]{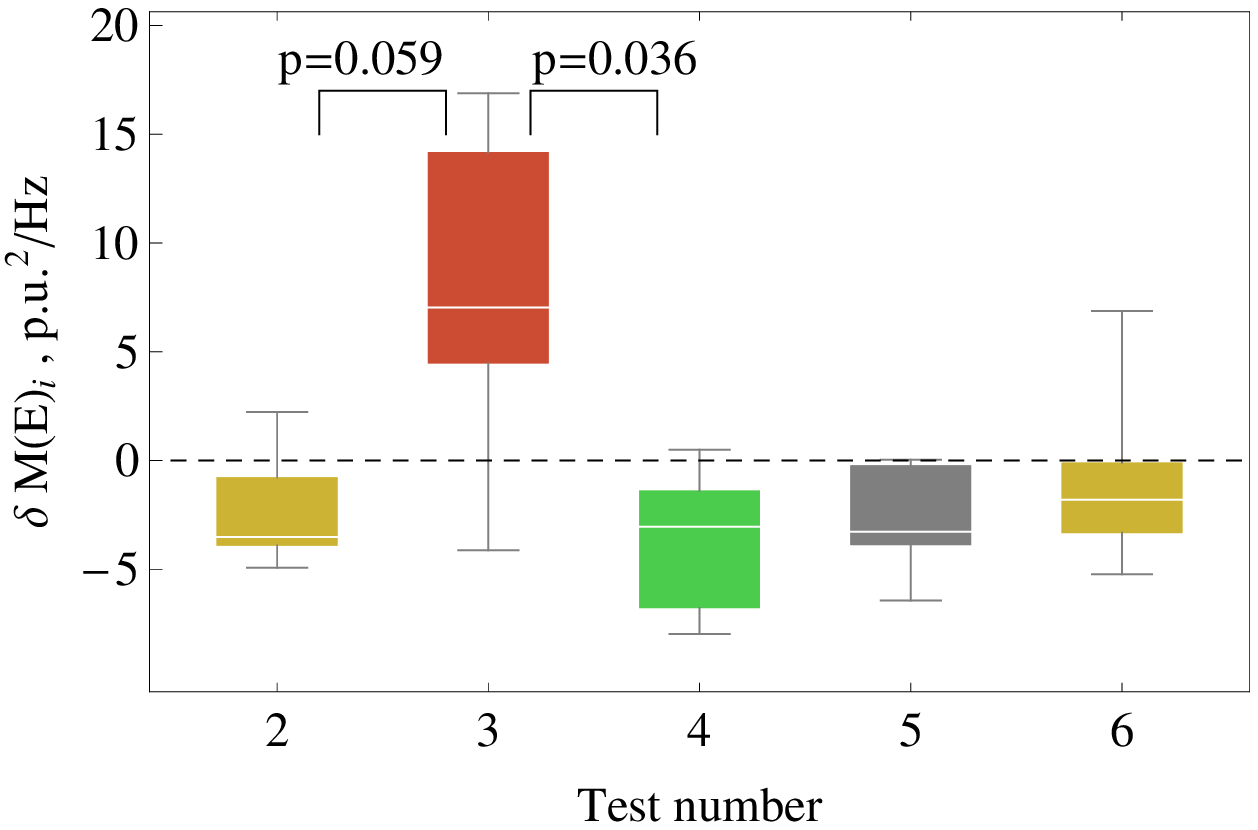}
\includegraphics[height=3.7 cm]{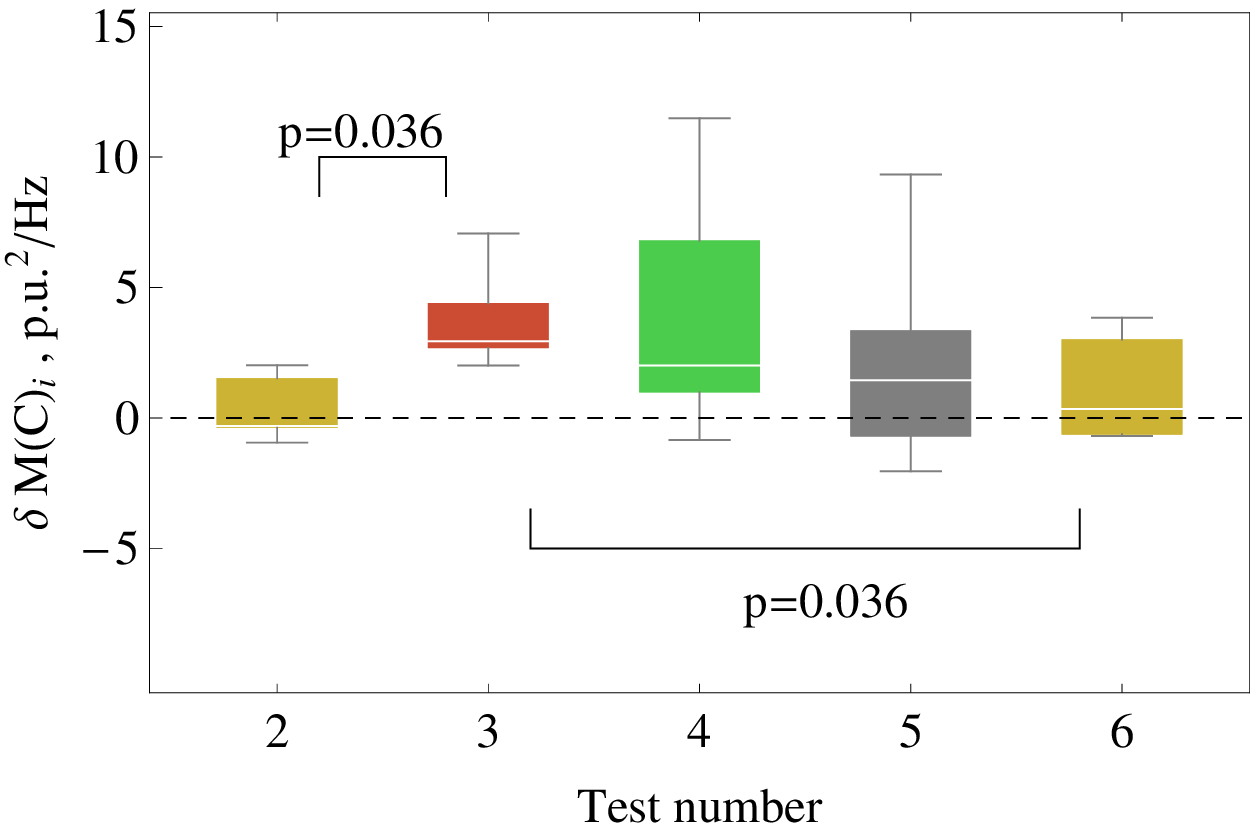}
\end{tabular}
\end{center}
\caption
{ \label{fig:EndoVar}
Variation of mean energy of endothelial associated (left panel) and cardiac associated (right panel) oscillations caused by skin loading}
\end{figure}

\begin{figure}
\begin{center}
\begin{tabular}{c}
\includegraphics[height=3.7 cm]{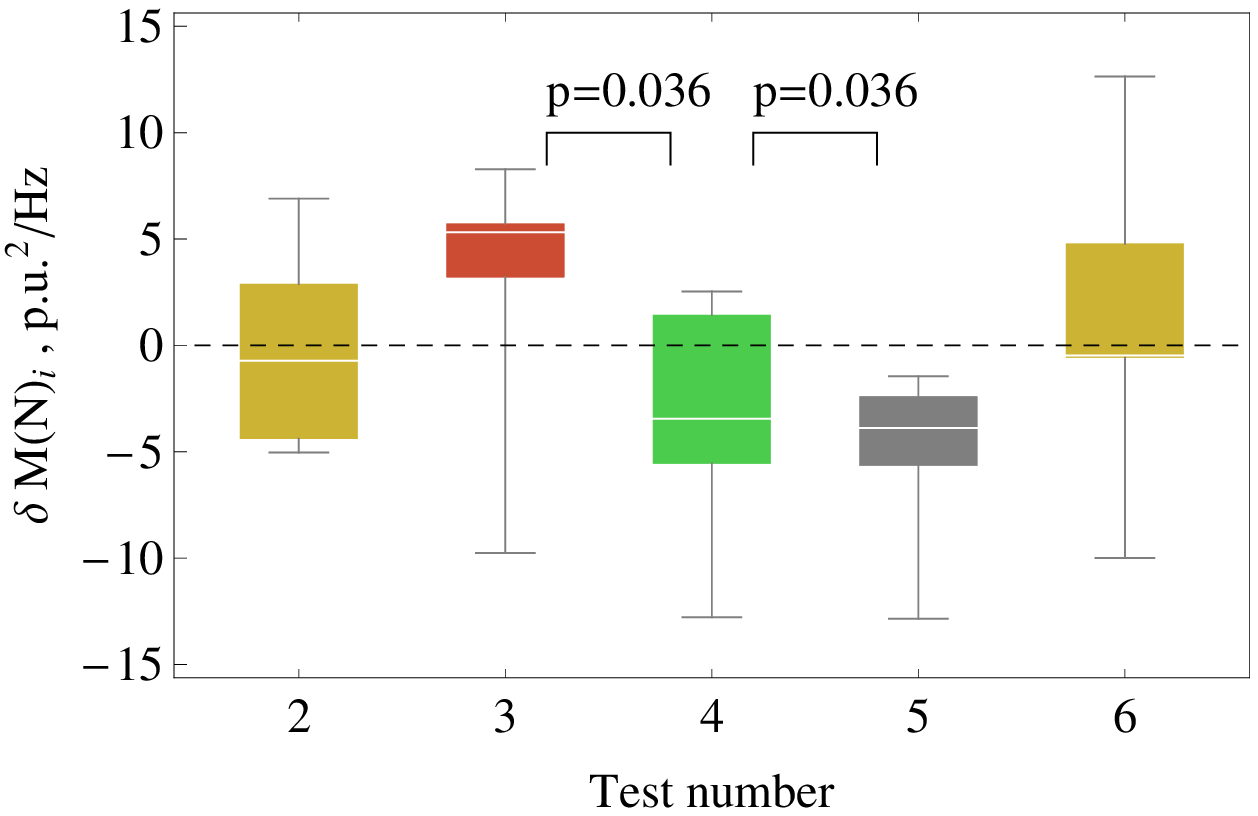}
\includegraphics[height=3.7 cm]{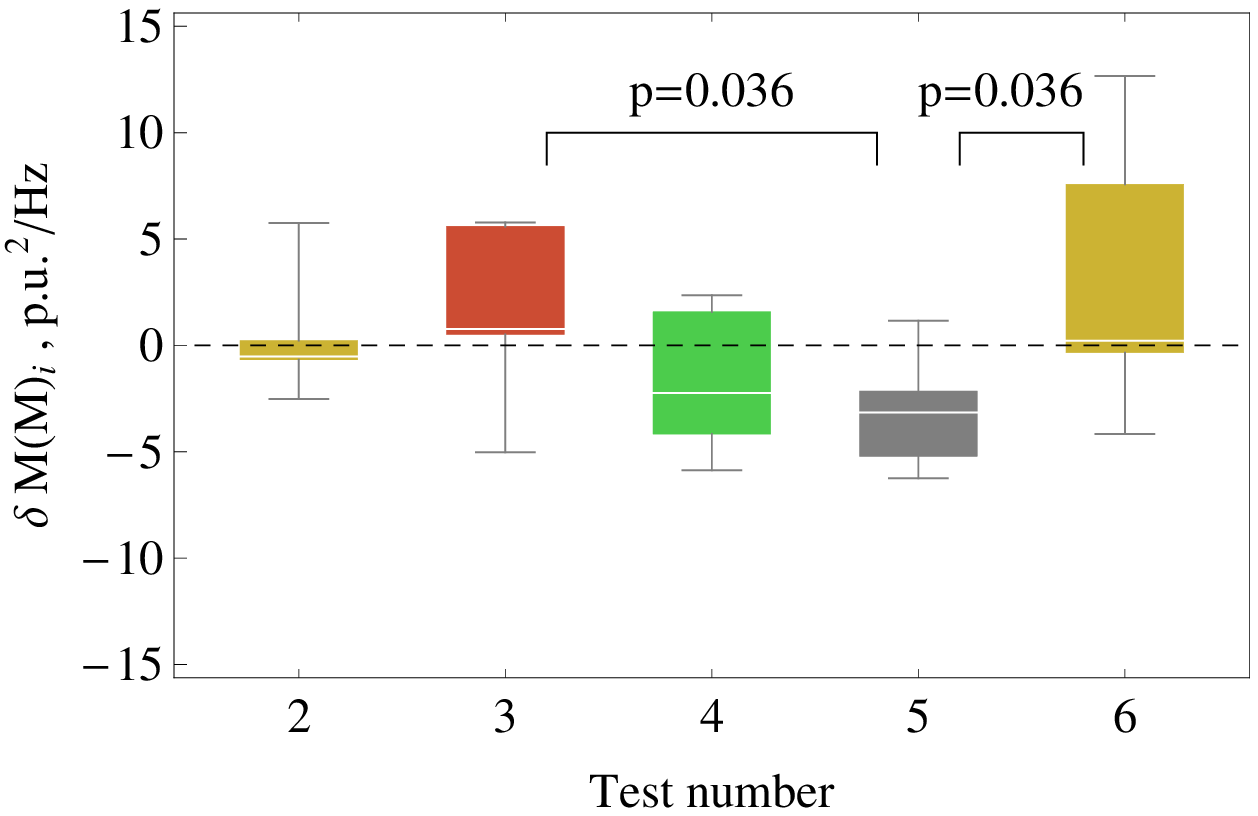}
\end{tabular}
\end{center}
\caption
{ \label{fig:NeuroVar}
Variation of the mean energy of neurogenic associated (left panel) and myogenic associated (right panel) oscillations caused by skin pressure }
\end{figure}

Like the  variation in the mean perfusion at gradually increasing local pressure, the oscillating components also vary (Fig.~\ref{fig:SpExamp}). We traced how the local loads that induce the variation of spectral energy in five frequency bands associated with different physiological mechanisms influence the vascular tone (Fig.~\ref{fig:EndoVar}-~\ref{fig:NeuroVar}). To this end, we used a relative variation in the mean spectral energy $M(\nu)_i-M(\nu)_1$ during tests 2..6 with respect to the basal conditions (test 1) for the frequency bands $\nu$ of endothelial ($E$), neurogenic ($N$), myogenic ($M$) and cardiac ($C$) oscillations. The respiratory frequency band was also studied, but significant differences were not found. To estimate these variations, we averaged energy in a frequency band for one subject and then plotted box-whisker diagram for all subjects.   

It has been found that the endothelium-associated pulsations during test 2 are weaker than the same oscillations in a basal state (left panel of Fig.~\ref{fig:EndoVar}). Further loading results in a significant increase in the low-frequency (0.01-0.02) pulsation. The  pressure exceeding 140 mmHg leads to a decrease in endothelial associated pulsations. This result was observed in all 30 records. After removing the leads from the loading tool (test 6), the level of pulsations restores up to the level of test 2. Note that the behavior of energy of the endothelial associated oscillations $M(E)_i$ is quite different from the behavior of the mean perfusion $P_i$. 

The dynamics of cardiac associated pulsations in peripheral vessels is slightly different. In contrast to the base level and weak loading ($M(C)_2$), an increase of $M(C)_3$ is observed in all samples during test 3. The energy of pulsations monotonically decreases and fully restores after removing the loading.

The evolution of myogenic and neurogenic associated oscillations is very similar. Moderate loading (90~mmHg) leads to an increase of amplitude of pulsations, which decreases and becomes lower than in basal state during tests 4 and 5. We observed the tendency to higher oscillations in these frequency bands after removing the loading (tests 2 and 6).
We have not found any statistically significant variations in respiratory oscillations of the blood flow on the periphery, that is why, such plot is not presented.

\section{Discussion}

Analysis of the influence of the local pressure test on the measured parameters of peripheral blood circulations has revealed a considerable variation in both mean and oscillation components of the LDF signal.

It has been found in this study that during the second stage of the test pressure induces vasodilation occur. Whereas increasing pressure  (stage 3), endothelial oscillations become significantly higher. Further pressure (stages 4 and 5) causes significant decrease in all frequency bands. 
 
Based on the obtained results, we assume that two mechanisms can be responsible for PIV and endothelial associated blood flow oscillations. 

On the one hand, it is known that low frequency pulsations are caused by the vasorelaxation effect of nitric oxide synthesized due to the shear stress on the internal vessel boundary. Decreased blood vessel lumen leads to higher blood velocity and thus higher shear stress \cite{Arnal1999, Michiels2003, Davies1995}. Previous studies \cite{Fromy2000} have demonstrated that blood vessels in microcirculation are susceptible to local vertical pressures acting as shear forces. The shear stress induces the capsaicin-sensitive nerve fibers to stimulate the endothelial release of vasodilatory nitric oxide, prostaglandins and acetylcholine, which causes relaxation of the smooth muscles of the vessels. Neuronal mechanosensitivity in the vasculature is believed to mediate PIV causing local hyperemia. Presumably this is a protective mechanism of autoregulation that can prevent tissue ischemia in response to an external non-noxious pressure application. So, this mechanism is one possible cause of increased blood flow at low pressure.

On the other hand, higher external pressure leads to blanch the skin tissues in the diagnostic volume \cite{Liasi2018}, namely clip capillaries. That is why we observe more vivid endothelial associated fluctuations formed in arterioles.

Oscillations belonging to other frequency bands stay stable during weak pressure (30 mmHg). Further loading (90 mmHg)  induces amplification of pulsations in all frequency bands. We assume that the pressure leads to the higher number of clamped  capillaries and thus to the higher input from arterioles in the LDF signal and, as a consequence, to  more vivid oscillations which are mainly formed in arterioles.

The use of the local pressure of 90 mmHg significantly increased the amplitude of myogenic associated oscillations in the LDF signal. The increase in myogenic activity can be a protective mechanism in response to the compression of the vessels by external pressure. Local pressure higher then 140 mmHg leads to a decrease of oscillating components. Removing the loading results in very rapid restoration of tissue perfusion. 

\section{Conclusion}

Local skin compression has an effect on the measured parameters of peripheral blood circulation. The obtained results led us to conclude that the LDF signal in mean and oscillation components varies considerably. Low pressure leads to PIV, so if a LDF diagnostic device uses a probe which weakly compresses the skin, the measured skin perfusion will be distorted and presumably overestimated. This should be taken into account when developing wearable LDF devices. 

The proposed protocol of local skin pressure allows one to evaluate PIV, as well as to trace the dynamics of blood flow oscillations. We have registered the most vivid variations of LDF during test stages 2, 3 and 4 or 5. To decrease the protocol duration, step 4 or 5 can be dropped from the consideration, because they give similar results. 

Also, researchers can use a shorter protocol of 40 minutes length combined of 10 minutes records for basal, 30, 90 and 140 (or 200) mmHg loadings. The moderate loading 90 mmHg allowed us to identify  spectral characteristics, which is of particular interest for those who are interesting in the oscillating dynamics of vasomotions. The proposed method (setup and protocol designs and a data processing algorithm) can be used to study the function of microcirculation and the endothelial function in microvessels especially under pathological conditions when other physiological tests are contraindicative.

\section*{Duality of Interest}
No potential conflicts of interest relevant to this article were reported.

\section*{Acknowledgment} 
This study was supported by the Russian Science Foundation under project N\textsuperscript{\underline{o}}18-15-00201.

\bibliographystyle{vancouver}
\bibliography{sample}

\end{document}